\documentclass{article}
\usepackage{spconf,amsmath,graphicx,tabulary}
\graphicspath{ {./images/} }

\usepackage{multirow}

\title{ENHANCEMENTS FOR AUDIO-ONLY DIARIZATION SYSTEMS}

\name{Dimitrios  Dimitriadis}

\address{Speech and Dialogue Research Group, Microsoft, Bellevue, WA\\
        \small{\tt{didimit@microsoft.com}}
    }
\begin{document}
\ninept
\maketitle
\begin{abstract}
In this paper two different approaches to enhance the performance of the most challenging component of a Speaker Diarization system are presented, i.e. the speaker clustering part. A processing step is proposed enhancing the input features with a temporal smoothing process combined with nonlinear filtering. We, also, propose improvements on the Deep Embedded Clustering (DEC) algorithm -- a nonlinear feature transformation. The performance of these enhancements is compared with  different clustering algorithms, such as the UISRNN, k-Means, Spectral clustering and x-Means. The evaluation is held on three different tasks, i.e. the AMI, DIHARD and an internal meeting transcription task. The proposed approaches assume a known number of speakers and time segmentations for the audio files. Since, we focus only on the clustering component of diarization for this work, the segmentation provided is assumed perfect. Finally, we present how supervision, in the form of given speaker profiles, can further improve the overall diarization performance. The proposed enhancements yield substantial relative improvements in all 3 tasks, with 20\% in AMI and 19\% better than the best diarization system for DIHARD task, when the number of speakers is known. 
\end{abstract}

\section{Introduction}
\label{sec:Intro}

Speaker diarization is the task of determining `who spoke when'~\cite{TrRe06} in a multi-speaker environment. It is an essential component for a variety of applications such as call center services, meeting transcriptions, etc,  thus the attention drawn from the research community~\cite{RKC2009, Anguera+_12}. Although, the process of diarization sounds inherently easy, it poses multiple challenges in practice, limiting its commercial deployment. 
One of the reasons is that the overall performance depends heavily on both the application scenario and the imposed constraints. For example, diarization focused on call center audio is  mostly about separating just two speakers, often in quite diverse acoustic environments. On the other hand diarization of meetings is much more challenging with multiple speakers, reverberation, back channeling, etc. As such, even determining  the number of active speakers still poses a  scientific challenge.

The main approach for the diarization systems is to first extract noise- and environment-invariant speaker embeddings and then cluster them. Most of the previous work is based on the i-vector embeddings~\cite{SDDG2014, Dehak+_2011, SDG2008, Dupuy+_2012}. However, the latest research work shows a shift from i-vectors to d-vectors~\cite{Variani+_2014, HMBS2016}, i.e. features based on the bottleneck layer output of a DNN trained for Speaker Recognition. Reasons for this shift are attributed to the enhanced performance, easier training  with more data~\cite{Zhang+_2018}, and  robustness  against speaker variability and acoustic conditions. The main difference between i-vectors and d-vectors are that the latter are extracted in a frame-level fashion from the bottleneck layer. 
%No assumptions about the d-vectors distribution are necessary, unlike the i-vectors, where it is assumed that they follow a Gaussian distribution.

In terms of clustering, most of the literature can be grouped into two approaches: the first, using supervised clustering~\cite{Zhang+_2018, YBB2018}, where the proposed system is trained on pre-segmented, speaker labeled utterances. These approaches usually provide a neural network modeling the speaker profiles on-the-fly and techniques like an HMM for the speaker sequence and segmentation. The second approach is by clustering the speakers in an unsupervised manner while the algorithm decides the data modality, i.e. the speaker number~\cite{YBB2018, DiFo17}. 

Along these lines,  the performance of the Deep Embedded Clustering (DEC) variations~\cite{XGF2016,GGLY2017} is also investigated and compared with the k-Means, Spectral Clustering (assuming the number of speakers is known). In the case of unknown number of speakers, we investigate the performance of the ``x-means'' algorithm~\cite{PeMo00}, where the number of speakers is determined on-the-fly.  To highlight the impact of the clustering performance, while the number of speakers is known, we will use oracle audio segmentations. We also compare the proposed approach against the fully-supervised UISRNN system.% where the diarization process becomes end-to-end.
%Finally, we investigate how knowing speaker profiles can improve the diarization performance.  

This paper is structured as follows: in Sec.~\ref{sec:Background}, a short overview of the d-vector extraction and the existing clustering approaches is provided. In Sec.~\ref{subsec:Feats}, the enhancements in the front-end are discussed, where the temporal filtering and the median are discussed.  The refinement in the DEC algorithm are discussed in Sec.~\ref{subsec:Improve_DEC}, where the loss function for training is revisited. In Sec.~\ref{sec:Experiments}, the experimental results on 3 different tasks, i.e. the AMI database~\cite{Carletta+_2005}, the DIHARD corpus~\cite{Sell+_2018}, and the internal meeting data, are presented. Finally, our latest findings are recapped and conclusions are presented in  Sec.~\ref{sec:conclusions}.

\section{Background}
\label{sec:Background}

\subsection{Clustering Methods}
\label{subsec:Clustering}

%\subsubsection{k-Means and Spectral Clustering}
%\label{subsubsec:kMeans}

The baseline clustering system is based on two of the most widely used clustering algorithms, i.e. the k-Means and the Spectral clustering. The k-Means clustering is an unsupervised  cluster analysis,  partitioning the samples into $k$ clusters, with $k$ known.
%, minimizing the total distance from their respective cluster mean values, called ``centroids''. Although, the problem is NP-hard, there are efficient heuristic algorithms converging quickly to a local optimum. 
Both, the k-Means and the expectation-maximization algorithms for GMMs are similar in a sense, i.e. using cluster centroids, i.e. `means', to model the data; however, k-means clustering finds clusters of comparable spatial variance, while the expectation-maximization mechanism allows clusters to have very different ones. The objective function is,
\begin{equation}
    \underset{S}{\arg \min} \sum_{i=1}^k{\sum_{\mathbf{x}\in S_i}{\|\mathbf{x}-\mathbf{\mu}_i \|^2}}
\label{eq:kMeans}
\end{equation}
where $\mathbf{x}$ are the samples, $\mathbf{\mu}_i$ are the centroids and $\mathbf{S}=\{S_1, S_2, \cdots,S_k\}$ the sets corresponding to the respective clusters. Before using the k-Means algorithm the input samples are whitened and their dimensionality is reduced using PCA. The $\mathbf{\mu}_i$ can be replaced by the speaker profiles, when initializing the process, incorporating prior knowledge to the system.

Spectral clustering is the default state-of-the-art unsupervised clustering algorithm for diarization providing very good performance. It is based on a similarity matrix $\mathrm{A}$ defined as a symmetric matrix, where $A_{i,j} \geq 0$ represents the similarity between any two data points with indices $i,\ j$. The spectral clustering approach employed here uses the k-Means on the eigenvectors of the `graph Laplacian' matrix of $\mathrm{A}$,~\cite{Luxb2007}. The intrinsic dimensionality reduction provides an additional robustness to the algorithm. 

Both algorithms have drawbacks, such as the random initialization step and the requirement for a preset number of clusters $k$, Eq.~\eqref{eq:kMeans}. The latter constraint is addressed with the `x-Means' algorithm~\cite{PeMo00} may start with a lower bound of $k$ and keep splitting the clusters until a stopping criterion is reached (or the purity of the clusters reaches a certain level). Herein, the Bayesian Information Criterion (BIC) is used~\cite{PeMo00}, 
\begin{equation}
    BIC(M_j)= I_j(D)-\frac{n_j}{2}\log N
\label{Eq:BIC}
\end{equation}
where $I_j(D)$ is the log-likelihood of the data according to the $j$-th model, and $n_j$ is the number of parameters in model $M_j$ (where $M_j$ the family of models) and $N$ the number of samples. The algorithm keeps splitting the clusters until all clusters are different enough based on BIC. A similar idea (albeit in a completely different approach) has been proposed in~\cite{Sell+_2018} while using Agglomerative Hierarchical Clustering.

\subsection{Deep Embedded Clustering -- DEC}
\label{subsec:DEC}

The motivation behind DEC is to transform the input features, herein speaker embeddings, to a space better separable in a given number of clusters. 
%The main challenges of the algorithm are due to the nature of the unsupervised training used. 
The clusters are iteratively refined based on a target distribution~\cite{XGF2016}. First, an autoencoder is trained while injecting noise, i.e. dropouts for the encoding part. The autoencoder learns a representation of the input features in a space of much lower dimensionality~\cite{XGF2016} (embeddings), while maintaining the separable properties of the features. The encoder outputs $z_i$ are used as input for the clustering component, iteratively refined by learning from their high confidence assignments. 
Specifically, the DEC model is trained with  the  KL-divergence between the $Q$ and $P$ distributions as the loss function, when matching the soft assignment $q_{ij}$ of the embedding $z_i$ to the cluster $j$ with the target distribution $p_{ij}$, Eq.~\eqref{Eq:DEC},
\begin{equation}
    L_c = KL(P\|Q) = \sum_i \sum_j p_{ij}\log\frac{p_{ij}}{q_{ij}}   
\label{eq:Loss1}
\end{equation}
and $q_{ij}\text{ and }p_{ij}$ are given by,
\begin{equation}
    q_{ij}=\frac{ \left(1+\| z_i-\mu_j\|^2/a\right)^{-\frac{a+1}{a}} }{\sum_l\left(1+\| z_i-\mu_l\|^2/a\right)^{-\frac{a+1}{a}}},\ \ p_{ij}=\frac{q^2_{ij}/f_i}{\sum_l q^2_{il}/f_l}
    \label{Eq:DEC}
\end{equation}
where $\mu_i$ the centroid of $i$-th cluster and $f_i$ is the soft cluster frequency with $f_i=\sum q_{ij}$. 
The DEC approach presents however some problems: first, the training of the autoencoder and the clustering steps are decoupled, according to Eq.~\eqref{eq:Loss1}. This is may lead to trivial solutions, especially when the encoded features are not discriminative enough, since there is no gradient back-propagation to the autoencoder. An initial fix was proposed in~\cite{GGLY2017}  adding a second loss term to preserve the local structure of the input features while improving their separability,
\begin{equation}
    L_r = \sum^n_{i=1} \| x_i-f(x_i)\| ^2
    \label{eq:Loss2}
\end{equation}
where $f(\cdot)$ is the encoder and decoder mappings combined. Now, the autoencoder is forced to  maintain local structure while improving discrimination, thus the features cannot collapse in the `trivial' space. 

However, the two-term loss function doesn't address another fundamental problem of the algorithm: there is no constraint for avoiding empty clusters, despite the fact that the number of classes is known.  The non-trivial solution might be implied in Eq.~\eqref{eq:Loss1} but it's not adequate. This issue can be further enhanced since the features are transformed to a low dimensionality space, according to the loss terms $L_c, \ L_r$, without constraints.

\subsection{Unbounded Interleaved-State RNN}

Lately, an online method called `Unbounded Interleaved-State Recurrent Neural
Network (UISRNN)' has been proposed in~\cite{Zhang+_2018} for fully supervised speaker diarization. The input to the algorithm is d-vectors and uses an RNN to keep track (as a different state) for the different speakers as they are interleaved in the time domain. The RNN is part of a Bayesian framework supporting an unknown number of speakers. Although there are fundamental differences with the other algorithms investigated here, we include results using this algorithm for comparison reasons.

\section{Enhancements of Speaker Clustering }
\label{sec:Enhancements}

\subsection{Improving Speaker Embeddings}
\label{subsec:Feats}

The d-vectors are extracted using a DNN~\footnote{A TDNN is used for the extraction, now called `x-vectors'~\cite{Sell+_2018}}, with stacked log-mel filterbank energy coefficients as input features. The output of the network is a one-hot speaker label~\cite{Zhang+_2017} (or equally the  probability of that particular speaker given the current input frame). The d-vectors are the output of the second to last DNN layer which is usually much shorter than the last one. This layer is called the `bottleneck' layer and its size depends on the implementation. 

The frame-based nature of the d-vectors leads to noisy frame estimates despite the very long input time-windows -- usually around $0.5\,sec$ or more of audio. Most approaches using d-vectors are aggregating them over the span of a segment by averaging. As such, the length of the input segments is one of the limitations for high-quality d-vectors with shorter segments corresponding to suboptimal clustering results, Sec.~\ref{sec:Experiments}.

Herein, we propose a different approach, where the d-vectors are first low-passed and then aggregated by a median filter, 
\begin{equation}
    \tilde{x}_n^i=x_n^i \ast F
\end{equation}
where $x_n^i$ is the $i$-th coefficient of the $n$-th frame and $F$ is a moving-average, FIR filter estimated as $F_j=F_{j-1}\ast F_0,\  j=1,\cdots N$, and $F_0[n]=\frac{1}{2}\left(\delta[n]+\delta[n-1]\right)$, where $\delta[\cdot]$ the Dirac function. This filtering process results in smoother, less noisy temporal trajectories of the d-vector coefficients. A median value for each segment is then extracted from these temporally smoothed vectors,
\begin{equation}
    \hat{x}^i = med(\tilde{x}^i_n),\ \ n\in[N_s,\ N_e]
\end{equation}
where $N_s,\ N_e$ the start/end segment frame index, respectively. The `smoothing and median filtering' approach has been found to outperform the widely used averaging scheme for several reasons. Assuming the d-vectors belonging to the same speaker are similar enough, the variations of adjacent d-vectors can be attributed to the phonetic content or the environmental noise and as such they can be discarded. Additional robustness is provided by the median filtering, where the outliers have smaller impact on the aggregated values compared to averaging.

\subsection{Improvements on Deep Embedded Clustering}
\label{subsec:Improve_DEC}

The second  contribution of the paper is revisiting the overall loss function and  adding a few algorithmic steps to the DEC algorithm. 

First, the possibility of empty clusters has to be addressed. The basic assumption of our approach is that the distribution of speaker turns is uniform across all speakers, i.e. all speakers contribute equally to the session. This assumption is not realistic in real meeting environments but it constrains the solution space enough to avoid the empty clusters without affecting overall performance.  
Under this assumption, the $n$ input samples are uniformly distributed over $k$ clusters, expressed by,
\begin{equation}
    L_u = KL(P\|U) = \sum_i \sum_j p_{ij}\log\cfrac{p_{ij}}{u_{ij}}
    \label{eq:Loss3}
\end{equation}
where $U$ is the uniform distribution or equally $u_{ij}=n/k$, the clusters are now forced to be more balanced, while penalizing clusters not following the uniform distribution.

An additional loss term  penalizes the distance from the centroids $\mu_i$. This MSE term is given by:
\begin{equation}
    L_{MSE}=\sum_{i\in S_j,j}\|x_i-\mu_j\|^2
    \label{eq:Loss4}
\end{equation}
This is similar to the k-Means criterion in Eq.~\eqref{eq:kMeans}, but it is now expressed as part of the loss-function. 

The loss function of the revisited DEC algorithm now becomes,
\begin{equation}
    L= \alpha L_c + \beta L_r + \gamma L_u + \delta L_{MSE}
    \label{eq:loss}
\end{equation}
Although not presented here, the weights $\alpha, \beta, \gamma \text{ and } \delta$ can be fine-tuned on some held-out data.

Finally, an additional  k-Means `re-calibration' step is included every few training iterations. 
The DEC algorithm uses the k-Means for initializing the centroids and then, it runs iteratively based on the loss functions.
Based on our experience, the $Q$ distribution, Eq.~\eqref{eq:Loss1}, can diverge from the target distribution and a reset is necessary. Such a reset on iteratively refined features ensures that the system cannot diverge to an `ill-conditioned' solution.

\section{Experiments}
\label{sec:Experiments}

\subsection{System Setup}
We  investigate the performance of the proposed components on 3 different tasks, the AMI~\cite{Carletta+_2005}, the DIHARD~\cite{Sell+_2018} and an internal meeting transcription task. The AMI dataset consists of 166 meetings with 4 speakers worth of 100h captured by multiple lapel mics and 2 microphone-arrays. We use only the lapel recordings (1 per speaker) for the segmentation and the mixed audio (the 4 channels are mixed into one) for diarization\footnote{The AMI dataset provides this audio signal after mixing all the lapel channels together}. Each set of speakers is used for 4 meetings. The DIHARD set is a collection of diverse recordings with a varying number of speakers, noise conditions and spoken languages. The task contains two tracks, with/without the transcriptions given. Herein, we utilize only the first track (with the known segmentations).
Finally, the third dataset contains two 1h-long internal meetings, i.e. `Meeting A' and `Meeting B'. There are 6 and 4 participants, respectively. The audio is recorded with a microphone array and processed by a fixed beamformer~\cite{Yoshioka+_19} keeping the top-beam, i.e. the most active in terms of signal energy. This single-channel audio is then processed for diarization. In all but the DIHARD task and for the case of x-Means in Table~\ref{tab:DIHARD}, the number of speakers and the segmentations are considered given. Any silence shorter than $0.25\,sec$, i.e. the collar, is treated as `speech' for training, testing and scoring. We use the ground-truth segmentations provided by the databases for the d-vector aggregation and the time boundaries are considered as potential speaker-turns. For the case of the internal meeting data, we use the segmentations provided by the Microsoft ASR decoder. Therefore, there are  (short) silence segments present in the ground-truth segmentations.
%The ground-truth segmentation is created by merging words into utterances (from the same speaker) when the in-between word silence is shorter than the collar.
Finally, since there is no overlapping speech detection\footnote{about 10\% of the speech is considered overlapping~\cite{YoDi+_19}.}, i.e. segments with more than one active speakers are assigned to the speaker already talking. Consequently, the diarization results appear worse in the sense that some segments with overlapping speech are simply ignored. 

The d-vectors are trained on $1M$ text-dependent utterances, i.e. wake-up phrases, with $5k$ speakers~\cite{Zhang+_2017}. The d-vector length, i.e. the bottleneck layer, is 200 coefficients long and takes as input $51$ frames of log-mel filterbank energy coefficients. There is no overlap between the speakers in the training set and the speakers in the test audio. 

We use PCA for dimensionality reduction of the input to clustering algorithms. The length for the resulting d-vectors is 70 coefficients. In the case of the AMI task, the same set of speakers is found in 4 separate sessions/meetings. Thus, the diarization output of the first meeting can be used to create speaker profiles, serving as initial centroids for the rest of the meetings, Eq.~\eqref{eq:kMeans}. Finally, we use the original features in the case of Spectral clustering and DEC, since it has an intrinsic dimensionality reduction process.

The autoencoder for the DEC algorithm has the following architecture with dense layer size of $200:2048:2048:15:2048:2048:200$. The loss function, Eq.~\eqref{eq:loss}, has the following weights $\alpha=0.1,\beta=1,\gamma=10,\text{ and } \delta=1$. The network is trained with the Adamax criterion with learning rate $l_r=0.001$ and batch size of 64. 

% ddim: Isos na prostheso merikes grammes edo..
The UISRNN model is trained on transcribed internal meeting data of about 100h. The number of speakers varies on the meeting. The frontend for the UISRNN system is the same d-vector network as described above. For the UISRNN, we use a beam of 6 and 3 passes to further refine the diarization results.

Finally, the scoring is held with the standard for diarization evaluations NIST tool~\cite{NIST04}.

\subsection{Results}
\label{subsec:Results}

First, we investigate how the length of the speech segments can affect the diarization performance when training the PCA transformations and estimating the speaker centroids. We present results in Tables~\ref{tab:Meeting}-\ref{tab:AMI}, where segments shorter than $0.5 \text{ to }1.5\,sec$ are ignored. The performance is measured as the ratio of segments (in sec) assigned to the right speaker.  As mentioned, `Meeting A' contains 6 speaker and  `Meeting B' only 4. We use k-Means for the clustering part and the metric is the `Clustering Recall', i.e. the ratio of correctly assigned speech over the available speech (in sec).

The diarization performance is improved as expected, Table~\ref{tab:Meeting}, by ignoring these short segments. However, this is not a viable solution since the shorter segments remain unassigned. The `temporal and median' filtering can improve the overall performance reaching close to the best possible performance, as shown in the last row of the table, without ignoring any segments. Shorter segments provide embeddings of lower quality. It is possible to greatly improve the diarization performance by distinguishing the segment processing according to their length. Also, the number of speakers, i.e. clusters, can affect performance. Diarization performance in `Meeting A' is worse than the corresponding on `Meeting b' because there are more speakers (6 speakers vs. 4). %In general, diarization with larger number of participants still remains a challenge, especially when speaker profiles, i.e. prior knowledge, are not available. 
\begin{table}[htb]
\centering
\caption{Meeting Task -- Clustering Recall (\%).}
\begin{tabulary}{\linewidth}{|L|c|c|}
 \hline
 \multicolumn{3}{|c|}{Meeting Task --  Clustering Recall (\%)}  \\ \hline\hline
                            & Meeting A & Meeting B            \\ \hline
UISRNN                       & 81.39 & 94.20 \\ \hline\hline
All Segments                 & 82.80 & 91.88 \\ \hline
Ignore segments $\leq 0.5s$  & 89.53 & 91.63 \\  \hline
Ignore segments $\leq 1.0s$  & 91.38 & 92.85 \\  \hline
Ignore segments $\leq 1.5s$  & 93.38 & 95.05 \\  \hline
Ignore segments $\geq 0.5s$  & 51.95 & 56.74 \\  \hline
All segments                 && \\ 
+ temp. filtering            & 88.20 & 92.52 \\ \hline
\hline
\end{tabulary}
\label{tab:Meeting}
\end{table}

The second experiment, in Table~\ref{tab:AMI}, investigates how using the longer segments, where the quality of the aggregated d-vectors is expected higher, to train our feature transformations and estimate the speaker centroids. In the case of UISRNN and DEC in Tables~\ref{tab:AMI} and~\ref{tab:DIHARD}, all available segments are used as input, whether the temporal filtering is applied or not. The results for the UISRNN algorithm are provided after setting the decoding beam to 6 and allowing the algorithm to iteratively refine the results with 3 passes. In order to make the comparisons fair, we ignore the part of the Diarization errors that correspond to the VAD functionality and we report only the recall of the system. Finally, the performance of both DEC versions, i.e. the original and the improved one, is presented. Results are reported for the raw and processed d-vectors.

As shown, the proposed pre-processing step for the d-vectors greatly improves performance with an additional rel. improvement of $22\%$ for the best performing algorithm, i.e. the `Improv. DEC' (last row of Table~\ref{tab:AMI}). 
The enhancement of the DEC algorithm yields an 6.5\% over the original DEC version and 19\% over the Spectral Clustering results (best baseline clustering algorithm). Note here that there is no need for embedding pre-processing. The DEC algorithm can keep the salient components based on the autoencoder.
Also, utilizing the longer segments for processing/training can make an impact on the overall performance. Difference around 19.3\% can be achieved by using these segments first.

\begin{table}[h!]
\centering
\caption{AMI Task -- Clustering Error in DER(\%). The results are averaged over the entire AMI database. The input audio signals are the mixed lapel channels.}
\begin{tabulary}{\linewidth}{|L|c|c|c|c|}
 \hline
 \multicolumn{5}{|c|}{AMI Task -- Clustering Error (\%)}  \\ \hline \hline
                & All Seg. & $\leq 0.5s$ & $\leq 1.0s$  & All Seg+Filt.          \\ \hline
UISRNN          & 12.52 &   N/A &   N/A  &  N/A\\ \hline
Orig. DEC       & 11.41 &   N/A &   N/A  & 12.43 \\ \hline\hline  
k-Means         & 17.70 & 16.32 & 13.56  & 16.44 \\ \hline
Spectral Cl.    & 13.52 & 13.41 & 10.65  & 13.20 \\ \hline
x-Means         & 19.37 & 17.90 & 13.69  & 17.69 \\ \hline 
Impr. DEC       & 10.66 & N/A   & N/A    & 11.87 \\  

%
%
%UISRNN          &   &   N/A & N/A  &  N/A\\ \hline
%Orig. DEC       & 12.43 & N/A   & N/A   & \\ \hline  
%k-Means         & 25.35 & 26.28 & 28.40 & 21.25\\ \hline
%Spectral Cl.    & 19.89 & 23.33 & 26.00 & 16.22\\ \hline
%x-Means         & 25.92 & 25.79 & 26.76 & 22.03\\ \hline 
%Impr. DEC       & 18.60 & N/A   & N/A   & 14.44\\  
%
\hline
\hline
\end{tabulary}
\label{tab:AMI}
\end{table}

For the DIHARD task, we have initialized the autoencoder for the DEC algorithm with the Devel set. However, the initialization of the autoencoder does not significantly affect the overall performance, i.e. we also tried  initializing the autoencoder on meeting data with similar results. We report results only on `Track 1' dataset since the use of VAD is beyond the scope of this paper. The input features are d-vectors pre-processed with the temporal filtering and median averaging, as in  Sec.~\ref{subsec:Feats}. The best published results for the DIHARD task can be found in~\cite{Sell+_2018}. 
\begin{table}[htb]
\centering
\caption{DIHARD Task. The number of speakers is considered known -- apart of the case of the x-Means algorithm. d-vectors with the temporal filtering are used for clustering.}
\begin{tabulary}{\linewidth}{|L|L|L|}
 \hline
 \multicolumn{3}{|c|}{DIHARD Clustering Errors -- DER(\%)}  \\ \hline \hline
                             & Devel & Eval            \\ \hline
State-of-the-art~\cite{Sell+_2018} & 18.17 & 23.99 \\ \hline\hline 
Original DEC            & 28.17 & 30.38 \\ \hline
k-means                 & 17.90 & 19.77 \\ \hline
Spectral                & 18.36 & 19.32 \\ \hline
x-means                 & 19.33 & 25.99 \\ \hline 
Improved DEC            & 18.69 & 21.40 \\ \hline
\hline
\end{tabulary}
\label{tab:DIHARD}
\end{table}

The results in Table~\ref{tab:DIHARD} show a relative improvement of $19.5\%$ over the state-of-the-art (SoA) results (for the Eval. set), when the number of speakers is known. However, the comparison is not entirely fair: the system in~\cite{Sell+_2018} learns in a supervised manner the best thresholds to determine the number of speakers. Herein, there is an advantage over the SoA system since herein the number of speakers is given. A more fair comparison would be by comparing the `x-Means' performance with the SoA system\footnote{This is also an unfair comparison because no fine-tuning of the x-means algorithm is done.}. Also,  `x-Means' results are 30\% worse than the 'k-Means' ones, this is mainly due to the unknown number of speakers.   
Determining how many active speakers are present is part of the future work.

\section{Conclusions}
\label{sec:conclusions}

In this paper, enhancements in two different components of a diarization system are proposed, i.e. the speaker embeddings and the clustering algorithms. Factors, like the  segment length, are also investigated how affect the overall performance. We show that diarization performance for segments only longer than $1.5s$ is 62\% relatively better (38.56\% for `Meeting B') than the case where all the segments are included. The proposed approach is able to recover around 31.4\% of the optimal performance while including all segments. Also, it is shown that it is better to use these long segments to train the clustering models and use them for the shorter ones.

Further, we show the proposed enhancement in the DEC algorithm can yield up to 31\% improvement in clustering performance. The additional terms in the loss function constrain the system to a smoother clustering behavior. The DEC algorithm is able to filter out non-relevant information via the bottleneck layer of the autoencoder.

Finally, it is shown that having a good estimate of the number of speakers is crucial for the overall performance of the system. Herein, we assume we know the number of speakers in advance thus, obtaining about a $20\%$ boost in performance for the DIHARD task, when comparing the k-Means with the x-Means algorithm.

\section{Acknowledgments}
\label{sec:acknows}

We would like to thank Jeremy Wong for his work on UISRNN during his internship with Microsoft. His work is the basis for all the UISRNN-related experiments.

\bibliographystyle{IEEEbib}
\bibliography{main}

\end{document}